\documentclass[aps,pra,amsmath,amssymb,amsfonts,showpacs,floatfix,onecolumn]{revtex4}

\usepackage{graphicx}
\usepackage{epsfig}
\usepackage[english]{babel}
\newcommand{\be}{\begin{equation}}
\newcommand{\bea}{\begin{eqnarray}}
\newcommand{\eea}{\end{eqnarray}}
\newcommand{\ee}{\end{equation}}

\def\one{\ensuremath{\hbox{$\mathrm I$\kern-.6em$\mathrm 1$}}}

\begin{document}

\title{Matrix product operator representations}

\author{V. Murg}
\affiliation{Max-Planck-Institut f\"ur Quantenoptik,
Hans-Kopfermann-Str. 1, D-85748 Garching, Germany.}

\author{J.I. Cirac}
\affiliation{Max-Planck-Institut f\"ur Quantenoptik,
Hans-Kopfermann-Str. 1, D-85748 Garching, Germany.}

\author{B. Pirvu}
\affiliation{Fakult\"at f\"ur Physik, Universit\"at Wien,
Boltzmanngasse 5, A-1090 Wien, Austria.}

\author{F. Verstraete}
\affiliation{Fakult\"at f\"ur Physik, Universit\"at Wien,
Boltzmanngasse 5, A-1090 Wien, Austria.}

\begin{abstract}
We show how to construct relevant families of matrix product
operators in one and higher dimensions. Those form the building
blocks for the numerical simulation methods based on matrix product
states and projected entangled pair states. In particular, we
construct translational invariant matrix product operators suitable
for time evolution, and show how such descriptions are possible for
Hamiltonians with long-range interactions. We show how those tools
can be exploited for constructing new algorithms for simulating
quantum spin systems.
\end{abstract}


 \maketitle

The study of strongly correlated quantum systems is currently
receiving a lot of attention. To a large extent, this is due to the
formidable progress that has been made in creating such systems
under controlled laboratory conditions such as in optical lattices
and ion traps. From the theoretical point of view, major new
insights have been obtained into characterizing the nature of the
wavefunctions associated to those strongly correlated systems. The
concept of matrix product states and its generalizations plays a
central role in those new insights, as it provides a sound
foundation and justification for the success of numerical
renormalization group methods and especially of DMRG
\cite{DMRG1,DMRG2}. Those insights have led to the development of
new algorithms for simulating quantum spin systems; most notable are
the algorithms for simulating time evolution
\cite{Vidal,TEBD,White,VGC} and the ones generalizing DMRG to higher
dimensions \cite{PEPS}.

In this work, we are concerned with the efficient construction of
so--called matrix product operators (MPO), the basic building blocks
for those novel algorithms. MPO were introduced in the paper
\cite{VGC,Zwolak} and form the operator analogue of matrix product
states. We will show how to construct translational invariant MPO's
in 1 and 2 dimensions that approximate real or imaginary time
evolution; in contrast to the TEBD/DMRG algorithms
\cite{TEBD,White}, the translational symmetry is not broken in the
Trotter step. This generalizes the constructions reported in
\cite{phase}. Second, we construct MPO descriptions for general
Hamiltonians with decaying long-range interactions. This is very
interesting in the light of simulating quantum spin systems with
long-range interactions.

Similar work for constructing MPO representations of Hamiltonians
has independently been reported in \cite{Crosswhite1,Mccullogh}.
Reference \cite{Mccullogh} gives a very nice presentation of matrix
product operators from the point of view of DMRG, and also contains
results on how to write spin chain Hamiltonians using MPO. Reference
\cite{Crosswhite1} explores the connection between matrix product
operators and Markov processes in depth, and also contains some
results on generalizations to higher dimensions. In reference
\cite{Crosswhite2}, an algorithm is devised to simulate quantum spin
chains with long-range interactions in the thermodynamic limit; it
also contains similar results as reported here on the approximation
of power law decaying interactions by sums of exponentials.

\section{Matrix Product Operator descriptions of exponentials}

\subsection{Construction}
Let us first start with a  simple example: suppose we want to
simulate real/imaginary time evolution under the Ising Hamiltonian
in transverse field

\[\mathcal{H}_{Is}=-\sum_{<ij>}\sigma_i^z\otimes\sigma_j^z
-B\sum_{i}\sigma^x_i\]

where only nearest neighbour interactions are considered; both the
one- and two-dimensional case will be considered. As usual, this
evolution will be approximated using a Trotter expansion, but we
want to do this is such a way that the translational invariance in
not broken. Therefore, we split the Hamiltonian in two parts
$\mathcal{H}=H_z+H_x$ where $H_z$ contains all terms with $\sigma^z$
operators and $H_x$ the ones with $\sigma^x$. Obviously, all terms
in $H_z$ commute, and therefore $O_z=\exp(\epsilon H_z)$ can be
calculated exactly. As we will show, $O_z$ has a very simple and
elegant MPO description, and of course $O_x$ has a trivial MPO
description as it is a product of strictly local operators. Time
evolution can now be described within the formalism of matrix
product states by evolving the MPS under the action of the MPO
$O_xO_z$.

Let us next show how the MPO of $O_z$ can be constructed. First,
observe that

\begin{eqnarray*} \exp(\epsilon Z\otimes Z)&=&\cosh(\epsilon)I\otimes
I+\sinh(\epsilon)Z\otimes Z\\
&=&\underbrace{\left(
     \begin{array}{cc}
       \sqrt{\cosh{\epsilon}} & 0 \\
     \end{array}
   \right)}_{B_0^T}
   \left(
     \begin{array}{c}
       \sqrt{\cosh{\epsilon}} \\
       0 \\
     \end{array}
   \right) I\otimes I+
\underbrace{\left(
     \begin{array}{cc}
       0 & \sqrt{\sinh{\epsilon}} \\
     \end{array}
   \right)}_{B_1^T}
   \left(
     \begin{array}{c}
       0\\
       \sqrt{\sinh{\epsilon}} \\
     \end{array}
   \right) Z\otimes Z\\
   &=&\sum_{ij} \left(B_i^TB_j\right)Z^i\otimes Z^j.
   \end{eqnarray*}
Here we used the notation $Z^0=I,Z^1=\sigma_z=Z$ and defined the
vectors $B_i$. Let us now consider the translational invariant 1-D
case of N spins with periodic boundary conditions
\begin{eqnarray*}
\exp\left(\epsilon\sum_i Z_iZ_{i+1}\right)&=&\prod_i\exp(\epsilon Z_iZ_{i+1})\\
&=&\sum_{i_1j_1i_2j_2...j_Nj_1}\left((B_{i_1}^TB_{i_2})(B_{j_2}^TB_{j_3})...(B_{j_N}^TB_{j_1})\right)Z_1^{i_1}Z_1^{j_1}\otimes
Z_2^{i_2}Z_2^{j_2}\otimes ...\\
&=&\sum_{i_1j_1i_2j_2...}{\rm
Tr}\left(B_{j_1}B_{i_1}^TB_{i_2}B_{j_2}^TB_{j_3}...B_{i_N}B_{j_N}^T\right)
Z_1^{i_1+j_1}\otimes
Z_2^{i_2+j_2}\otimes ...\\
&=&\sum_{k_1k_2...}{\rm
Tr}\left(\underbrace{\left(\sum_{i_1}B_{i_1\oplus
k_1}B_{i_1}^T\right)}_{C^{k_1}}\left(\sum_{i_2}B_{i_2\oplus
k_2}B_{i_1}^T\right)...\right)Z_1^{k_1}\otimes
Z_2^{k_2}... \\
&=&\sum_{k_1k_2...}{\rm
Tr}\left(C^{k_1}C^{k_2}...C^{k_N}\right)Z_1^{k_1}\otimes
Z_2^{k_2}... \\
\end{eqnarray*}
In the third step, we made use of the cyclic property of the trace,
and in the fourth step, we made a change of variables $k_1=i_1\oplus
j_1$ where binary arithmetic is assumed. We have therefore proven
that $\exp\left(\epsilon\sum_i Z_iZ_{i+1}\right)$ has a very
efficient matrix product description with the matrices $C^k$ given
by

\begin{eqnarray*}
C^0&=&\sum_{i}B_{i}B_{i}^T=\left(
                                          \begin{array}{cc}
                                            \cosh(\epsilon) & 0 \\
                                            0 & \sinh(\epsilon) \\
                                          \end{array}
                                        \right)\\
C^1&=&\sum_{i}B_{i\oplus 1}B_{i}^T=\left(
                                          \begin{array}{cc}
                                            0 & \sqrt{\sinh(\epsilon)\cosh(\epsilon)} \\
                                            \sqrt{\sinh(\epsilon)\cosh(\epsilon)} & 0 \\
                                          \end{array}
                                        \right)\\
\end{eqnarray*}
A big advantage of this precise MPO formulation is that is is
symmetric; the spectral properties of the associated transfer
operator are hence well behaved, which is important if used in
algorithms with periodic boundary conditions \cite{inpreparation}.

In two dimensions, we can repeat exactly the same argument and
obtain the PEPS description of the operator

\[\exp\left(\epsilon\sum_{<ij>}
Z_iZ_j\right)=\sum_{x_1x_x...}F(C^{x_1},C^{x_2},...)Z^{x_1}_1\otimes
Z^{x_2}_2\otimes ...
\] with tensors
\[
C^x_{\alpha\beta\gamma\delta}=\sum_{
i+j+k+l=x}B_{i}(\alpha)B_{j}(\beta)B_{k}(\gamma)B_{l}(\delta).
\]
Here $B_i(\alpha)$ means the $\alpha$ component of the vector $B_i$,
$x\in\{0,1\}$ and the sum is taken over $i,j,k,l=0:1$ with the
condition that $i+j+k+l=x$ in binary arithmetic. This proves that
the PEPS description of the operator $\exp\left(\epsilon\sum_{<ij>}
Z_iZ_j\right)$ has bond dimension 2. Note that no approximations
were made and as such this statement is valid for any value of
$\epsilon$. In particular, this gives the MPO description for the
classical Ising partition function; its free energy can therefore be
calculate by contracting the tensor network consisting of tensors
$C^0$.

The previous analysis can trivially be generalized to the case of
any Hamiltonian that is a sum of commuting terms: for this class of
Hamiltonians, $\exp(\epsilon H)$ has a very simple matrix product
operator description. As this holds for any $\epsilon$, it also
holds for all thermal states,  and by taking $\epsilon\rightarrow
-\infty$ it is proven that all ground states of such Hamiltonians
have exact MPO descriptions that can easily be constructed. Notable
examples of this is the toric code state of Kitaev and the family of
string net states \cite{Kitaev,LevinWen}.

From numerical considerations, it is useful if the matrices/tensors
occurring in the MPO description are real and symmetric. There are
some tricks of how to achieve this. Consider for example the
Heisenberg antiferromagnetic Hamiltonian

\[\mathcal{H}_{Heis}=\sum_{<ij>}\left(X_iX_j+Y_iY_j+Z_iZ_j\right).\]

The operator $\exp\left(-\beta\mathcal{H}_{Heis}\right)$ can be
decomposed in Trotter steps consisting of $H_x,H_y,H_z$, and every
Trotter term involves operators of the form $\exp(-\epsilon H_x)$.
As we saw in the previous section, the associated matrices involve
terms like $\sqrt{\sinh(\epsilon)}$, which becomes complex when
$\epsilon>0$. What we can do however is a change of basis on every
second site (this obviously only works for bipartite lattices),
where we rotate the spins with the unitary operator $Y=\sigma_y$;
this maps $X_{2n}\rightarrow -X_{2n},Y_{2n}\rightarrow Y_{2n},
Z_{2n}\rightarrow -Z_{2n}$. On the level of the Hamiltonian, this
flips the sign of the $H_x$ and $H_z$ interactions, for which the
associated operators $\exp(+\epsilon H_x)$ have indeed real and
symmetric MPO descriptions. The problem seems to remain however with
the operator $\exp(-\epsilon H_y)$. This can however easily be cured
by defining the real antisymmetric matrix $\tilde{Y}=iY$ for which
$H_{\tilde{y}}=-H_y$ when we replace all operators $Y$ by
$\tilde{Y}$. Next, $\exp(+\epsilon H_{\tilde{y}})$ can again be
expressed as a MPO; however, we have to be careful as
$\tilde{Y}.\tilde{Y}=-I$ as opposed to $+I$. Looking back at the
derivation of the MPO for the Ising case, we can easily see that
this sign can be absorbed into $C$, and we can express \[
\exp\left(-\epsilon\sum_{i} Y_iY_{i+1}\right)=\sum_{k_1k_2...}{\rm
Tr}\left(\tilde{C}^{k_1}\tilde{C}^{k_2}...\right)\tilde{Y}_1^{k_1}\otimes
\tilde{Y}_2^{k_2}...\]
 as a MPO with matrices
 \begin{eqnarray*}
\tilde{C}^0&=&\sum_{i}B_{i}B_{i}^T.(-1)^i=\left(
                                          \begin{array}{cc}
                                            \cosh(\epsilon) & 0 \\
                                            0 & -\sinh(\epsilon) \\
                                          \end{array}
                                        \right)\\
\tilde{C}^1&=&C^1=\sum_{i}B_{i+1}B_{i}^T=\left(
                                          \begin{array}{cc}
                                            0 & \sqrt{\sinh(\epsilon)\cosh(\epsilon)} \\
                                            \sqrt{\sinh(\epsilon)\cosh(\epsilon)} & 0 \\
                                          \end{array}
                                        \right)\\
\end{eqnarray*}

Of course the same can be done in two dimensions. Here we get

\[\tilde{C}^x_{\alpha\beta\gamma\delta}=\sum_{
i+j+k+l=x}B_{i}(\alpha)B_{j}(\beta)B_{k}(\gamma)B_{l}(\delta).\sqrt{-1}^{(x+i+j+k+l)}\]
where the sum in the power $\sqrt{-1}^{(x+i+j+k+l)}$ is \emph{not}
in binary arithmetic. This clearly leads to a real translational
invariant MPO parametrization.

\subsection{Algorithms}

It is now obvious how to turn those MPO-descriptions to our
advantage for constructing new algorithms for the simulation of
quantum spin chains.

Let us first consider the case of imaginary time evolution, where
the goal is to evolve a state in imaginary time such as to simulate
a thermal (finite $\beta$) or ground state
($\beta\rightarrow\infty$). Obviously, we will use the
Trotterization described in the previous section. The big advantage
there is that the translational invariance is never broken, and
furthermore that the matrices involved in the MPS description of the
MPO are real and symmetric. In particular, that means that, if we
start with a translational invariant MPS with real symmetric MPS
description, then it will stay like that during the whole course of
the evolution. This has a dramatic effect on the numerical
conditioning and stability of the algorithm.

The algorithm for time evolution is now as follows: given a
translational invariant MPS with matrices $\{A^i\}$ with bond
dimension $D$ and MPO with matrices $\{B^i\},\{X^i\}$ of dimension
$D'$, we want to find a way of representing cutting the bond
dimension of the MPS $\{C^i\}$ given by

\[C^i=\sum_{jk}A^j\otimes B^k \langle i|X^k|j\rangle\]

in an optimal way. This can easily be done as follows: calculate the
leading eigenvector $x$ of the transfer operator $E=\sum_i
C^i\otimes C^i$ (note that $E$ is symmetric and as such this is a
very well conditioned problem). Rewriting $x$ as a $DD'\times DD'$
positive semidefinite matrix, we can easily calculate its singular
value decomposition $x=U\Sigma U^\dagger$. We now define the
projector/isometry $P$ as the rectangular matrix consisting of the
first $D$ rows of $U$, and act with this $P$ on the matrices $C^i$.
The updates matrices $A^i$ are therefore obtained by $A^i\equiv
P^\dagger C^i P$ which is obviously still symmetric and real.
Clearly, all those steps have to be done in such a way as to exploit
the sparse nature of the problem, such as done in DMRG, which leads
to a complexity that scales like $D^3$. Also, if the eigenvalues
that are thrown away are not small enough, we can always increase
the bond dimension.

The big advantage of this procedure is that it is extremely well
conditioned and very efficient to implement. This allows to go to
very large bond dimensions. Notably, as compared to the $TEBD$
algorithms, we do not have to take inverses at any time (because the
gauge degrees of freedom are trivial as they consist of unitary
matrices), and furthermore it works equally well if the MPO is very
far from the identity operator (this is important in the context of
PEPS algorithms).

The same ideas can of course be used in the case of real time
evolution. In that case, the matrices involved become complex
symmetric, and it might be benefitial  to apply some gauge
conditions to optimize the stability. This can be done as follows:
given $x$, we want to find the complex (symplectic) matrix $Q,
Q.Q^T=\openone$ such that the condition number (i.e. smallest
divided by largest singular value) of the matrix $QxQ^\dagger$ is as
large as possible. This optimization problem can be solved
recursively as follows: calculate the singular value decomposition
of $x=vsv^\dagger$, choose the generator $G_k=-G_k^T$ as $G_k={\rm
Im}(v_Dv_D^\dagger-v_1v_1^\dagger)$, and make the substitution
$x\rightarrow \exp(-i\epsilon G)x\exp(i\epsilon G)$ for small enough
$\epsilon$, and repeat this until convergence. Convergence is
equivalent to the derivative of the condition number being zero. The
final gauge transform to be implemented is the product of all
infinitesimal transformations $Q=\prod\exp(i\epsilon G_k)$. Note
again that all of this becomes trivial in the case of real symmetric
matrices (such as occurring in imaginary time evolution): in that
case the $Q$ cannot change the condition number as they are unitary.

We have tested those new algorithms on the critical Ising and
Heisenberg spin chain models, and obtained results that are
consistent with what we expected. In particular, for the Heisenberg
antiferromagnetic spin chain, we obtain a precision of
$(E_{D=64}-E_{exact})/E_{exact}=2.83*10^{-6}$ with very modest
calculations. In the case of the critical Ising in transverse field,
we get $(E_{D=64}-E_{exact})/E_{exact}=1,10*10^-9$.

The algorithms for the 2-D analogue will be discussed elsewhere
\cite{inpreparation}.

\section{Matrix Product Operator descriptions of Hamiltonians with
long-range interactions}

\subsection{Construction}

Let us next investigate how to represent Hamiltonians with
long-range interactions of the form

\[\mathcal{H}=\sum_{ij}f(i-j) Z_iZ_j\]

with $f(i-j)$ some decaying function. The first question to ask is
whether it is still possible to find an exact MPO description of
$\hat{O}=\exp(\epsilon \mathcal{H})$. It can easily be seen that
this is not possible if the function $f(x)$ does not vanish at some
finite distance: otherwise, the action of $\hat{O}$  on a MPS could
increase the Schmidt number over any cut with an arbitrary large
amount, and hence no finite MPO description is possible. This is the
reason why the transfer matrix approach in classical 1-D spin
systems breaks down for such long-range interactions.

So let's be less ambitious and try to find a MPO description of the
Hamiltonian itself. This is interesting for several reasons: first,
this is useful in constructing algorithms for time evolution using
iterative methods like Lanczos, and second, it allows to calculate
quantities like $\langle\psi|H^2|\psi\rangle$ efficiently.

As a start, let us consider a general 1-D spin 1/2 Hamiltonian with
nearest neighbour interactions. If the Hamiltonian is translational
and reflection invariant, then there always exists a basis such that
the Hamiltonian can be written as
\begin{displaymath}
\mathcal{H}=\sum_{\alpha,i} \mu_\alpha\sigma_\alpha^i\otimes
\sigma_\alpha^{i+1} +\sum_j \hat{O}^j
\end{displaymath}
where $\hat{O}$ can be any one-qubit operator. Similarly to the
construction of MPS descriptions of the W-state
\cite{MPSgeneration}, a MPO can be constructed to represent this
$\mathcal{H}$ by making use of nilpotent matrices:

\begin{eqnarray*}\mathcal{H}&=&\sum_{i_1i_2...}\left(v_l^TB_{i_1}B_{i_2}...B_{i_N}v_r\right)X_{i_1}\otimes X_{i_2}\otimes ...X_{i_N}\\
X_0&=&I\hspace{1cm} X_1=\sigma_x \hspace{1cm} X_2=\sigma_y \hspace{1cm} X_3=\sigma_z \hspace{1cm} X_4=\hat{O}\\
v_l&=&|0\rangle\hspace{2cm} v_r=|4\rangle\\
B_0&=&|0\rangle\langle 0|+|4\rangle\langle 4|\\
B_1&=&|0\rangle\langle 1|+\mu_1|1\rangle\langle
4|\hspace{1cm}B_2=|0\rangle\langle 2|+\mu_2|2\rangle\langle
4|\hspace{1cm} B_3=|0\rangle\langle 3|+\mu_3|3\rangle\langle
4|\\
B_4&=&|0\rangle\langle 4|
\end{eqnarray*}
The simplest way of deriving this is to think about the Hamiltonian
a Markov process with 5 possible symbols (remember that MPS can be
constructed using Markov processes), such that a symbol
$X_1,X_2,X_3$ is always followed by itself and then all zeros $X_0$,
and $X_4$ by all zeros. As such, one can easily prove that $D=5$ is
optimal in this case because this is the operator Schmidt number of
the Hamiltonian when splitting it into two pieces. Note that if only
Ising interactions would have been considered, then $D=3$ would have
been sufficient and we could have chosen
\[ B_0=|0\rangle\langle 0|+|2\rangle\langle 2|\hspace{1cm}
B_1=|0\rangle\langle 1|+\mu_1|1\rangle\langle 2|.\] Note that there
is no need for $B_2,B_3,B_4$ in that case.

It is obvious how to generalize this description to the case of
higher dimensional systems and to the case of exponentially decaying
interactions. Let us first look at the case of exponentially
decaying interactions. By adding diagonal terms to $B_0$
\[B_0=|0\rangle\langle 0|+\lambda_x|1\rangle\langle 1|+\lambda_y|2\rangle\langle 2|+\lambda_z|3\rangle\langle 3|+|4\rangle\langle
4|\] we can immediately check that that the corresponding
Hamiltonian / MPO is given by
\[\mathcal{H}=\sum_{\alpha,i<j}
\mu_\alpha\lambda_\alpha^{i-j}\sigma_\alpha^i\otimes
\sigma_\alpha^{j} +\sum_j \hat{O}^j\] which is a spin chain with
exponentially decaying interactions.

Unfortunately, it is impossible to get exact MPO descriptions when
the interactions are decaying following a power law. However,
inverse polynomials can pretty well be approximated by sums of
exponentials (this is the reason why DMRG is able to reproduce the
correlations in critical models pretty well). Hamiltonians with
power law decay of correlations can therefore be well approximated
by sums of MPO's, which is itself a MPO. Actually, very few
exponentials are needed to get a good approximation, even at large
distances. The problem of finding the optimal weights and exponents
for such an approximation problem for a general function $f(k)$,
i.e.
\[\min_{x_i,\lambda_i} \sum_{k=1}^N|f(k)-\sum_{i=1}^n
x_i\lambda_i^k|,\] is not completely trivial. In the appendix, we
present a simple method that solves this optimization problem for
general $f(k)$ and a given number of exponentials $n$ and a number
of sites $N>n$ (the method works for any functions, and returns
complex exponents in the case of oscillating functions as should
be). If we choose power law decay with cube power 3, N=1000 and n=10
then the above cost function is $10^{-5}$ (the maximal difference
between the function and the approximated one is $5.10^{-8}$). This
maximal difference falls to $3.10^{-6}$ for power 2 and $3.10^{-4}$
for power 1.

In conclusion, we found the exact MPO description for Hamiltonians
with exponentially decaying interactions. Hamiltonians with power
law decay can be approximated very well using sums of such MPO. The
matrix product operators obtained for the description of
Hamiltonians are of a very different form than the ones obtained by
taking the exponential. The main difference is that the
corresponding transfer matrices will always contain a Jordan block
structure, and one has to be careful in dealing with such situations
when considering the thermodynamic limit.

Let us now turn to the 2-dimensional case. Let us again first
consider the square lattice with only nearest neighbour
interactions. There is a very simple way of writing down a PEPS
description that achieves the task: first, consider the MPS
\[|W\rangle=\sum_{k}|0\rangle_1|0\rangle..|0\rangle|1\rangle_k|0\rangle...|0\rangle_{N^2}\] which is the equal
superposition of having one spin up and all other ones down over all
sites. Note that this MPS has bond dimension 2, and can therefore
trivially be represented as a PEPS with bond dimension 2. The idea
is that this excitation specifies where to put an interaction. Let
us next consider the tensors

\begin{eqnarray*}
B^0_{i;\alpha,\beta,\gamma,\delta}&=&|0\rangle\langle 0000|\\
B^1_{i;\alpha,\beta,\gamma,\delta}&=&|1\rangle\langle
00|\left(\langle 01|+\langle 10|\right) + |0\rangle \left(\langle
01|+\langle 10|\right)\langle 00|
\end{eqnarray*}
where we assume that the indices $\alpha,\beta$ are the left
respectively top indices,  and the associated operators

\begin{eqnarray*}
X^0&=&I\\
X^1&=&Z
\end{eqnarray*}

It can readily be seen that we get the Ising Hamiltonian if we act
with the $|W\rangle$ state on the fifth index of the tensors: the
$|W\rangle$ state puts one index $i$ equal to one, and the other
terms are such that an interaction to the right and below it will be
created. The total bond dimension of the corresponding PEPS
(including the $|W\rangle$) is therefore $4$.

Decaying interactions between one spin and all other ones can
however be obtained in a much more elegant way; as we will show, it
is even possible to model power law decay of interactions exactly.
The idea is as follows: the critical classical Ising model in 2
dimensions has power law decay of correlations. Consider the quantum
state
\[|\psi_\beta\rangle=\exp\left(-\beta\sum_{<ij>} Z_iZ_j\right)\left(|+\rangle\right)^{\otimes N} \] where
$|+\rangle$ stands for the superposition $|0\rangle+|1\rangle$. This
is obviously a PEPS, as it is obtained by acting with a MPO (see
earlier) on a product state. The partition function of the Ising
model at temperature $\beta$ is obtained by calculating the overlap
\[(\langle +|)^{\otimes N}|\psi(\beta)\rangle,\] and correlation
functions between  two spins are obtained by replacing the
corresponding $\langle +|$ at the left side of this expression by
$\langle -|=\langle 0|-\langle 1|$. Instead of the $|W\rangle$ state
in the previous example, we will use a state $|W"\rangle$ that is
the equal superposition of two excited spins as opposed to one. The
MPS description of $|W"\rangle$ has bond dimension 3 and is similar
to the ones derived for the 1-D Hamiltonians with exponential decay
where we put the parameter $\lambda=1$ (i.e.$B_0=I$). This gives us
all the necessary ingredients to construct the MPO description of
the Hamiltonian:

\begin{eqnarray*}
\mathcal{H}&=&\sum_{i_1i_2...i_N}\left(\langle x^{i_1}|\langle
x^{i_2}|...\langle
x^{i_N}|\right)\left(|\psi_\beta\rangle|W"\rangle\right)
X^{i_1}\otimes X^{i_2}....\otimes X^{i_N}\\
|x^0\rangle&=&|0\rangle|+\rangle, \hspace{2cm}
|x^1\rangle=|1\rangle|-\rangle\hspace{2cm} X^0=I\hspace{2cm} X^1=Z
\end{eqnarray*}

Here the vectors $|x^i\rangle$ act on two qubits, one on the
corresponding qubit of $|\psi_\beta\rangle$ and the other one on
$|W"\rangle$. The $|W"\rangle$ state enforces that exactly two
operators $X^i$ will be nontrivial, and $|\psi_\beta\rangle$ gives
the right weight to the associated interaction. As those are
products of PEPS, the result is a PEPS with bond dimension $2\times
3=6$. This is am amazing result: as opposed to the 1-D case, there
is an exact PEPS description for Hamiltonians with 2-body
interactions that decay as the $r^{-\nu}$ with $\nu=1$ the critical
exponent of the Ising model.

Obviously, this construction can be repeated for any classical spin
model, and hence many different exponents can in principle be taken.
It is as yet an open question how to engineer the PEPS such as to
get a specific exponent, although very good approximations can again
be obtained by making use of sums of exponentials.

\subsection{Algorithms}

It is obvious how to make use of all this in algorithms for
simulating quantum spin chains. First of all, it is clear how to
extend the variational MPS method described in \cite{Porras} to the
present case. For this, we have to consider finite systems with open
boundary conditions and matrix product states that have
site-dependent matrices in their MPS description. The optimization
$\langle\psi|H|\psi\rangle$ can then be done using the alternating
least squares method described in \cite{Porras}. As expected,
numerical tests showed very good convergence properties.

The problem of time evolution is a little bit more challenging, as
we cannot use the Trotterization tricks. However, Krylov-based
methods can of course be used (see \cite{Garcia} for a review), and
are the method of choice here.

These finite dimensional algorithms can not readily be generalized
to the infinite case however. A sensible way for determining ground
state energies in that limit would be to first use the finite
dimensional algorithm just described, and then use a brute-force
gradient-based optimization method for optimizing the infinite case.
For this we need to be able to calculate expectation values
$\langle\psi|\hat{O}|\psi\rangle$ in the thermodynamic limit, i.e.
when $|\psi\rangle$ is an infinite MPS with matrices $\{A^i\}$ and
$\hat{O}$ the MPO description of a Hamiltonian with exponentially
decaying interactions. The idea is to consider a family of MPO $O_N$
whose support is limited to  $N$ sites (i.e. the Hamiltonian does
only act on $N$ sites), calculate the energy with respect to the
infinite MPS (this energy will scale linearly in $N$), and the take
the thermodynamic limit to calculate the energy per site. It holds
that
\[\langle\psi|\hat{O}_N|\psi\rangle=\langle L|E_H^N|R\rangle\]
where
$|L\rangle=|x_l\rangle|v_l\rangle$,$|R\rangle=|x_r\rangle|v_r\rangle$
with $|x_l\rangle,|x_r\rangle$ the left/right eigenvectors
 of the transfer matrix $E_0=\sum_i
A_i\otimes \bar{A}_i$;  the vectors $v_l,v_r$ are the ones used in
the MPO description of the Hamiltonian, and

\[E_H=\sum_{ijk}A_i\otimes B_j\otimes \bar{A}_k \langle
k|X_j|i\rangle.\]

The eigenstructure of $E_H$ is nontrivial because it has Jordan
blocks. In the present case, the only relevant blocks are of size
$2$ (larger blocks would lead to an energy that scales superlinearly
with the size of the support of $H$, which cannot be). Using the
notation used before, one sees that the left/right eigenvector
corresponding to the largest eigenvalue in magnitude $d_0$ is given
by $\langle q_l|=\langle x_l|\langle 0|$ /
$|q_r\rangle=|x_r\rangle|0\rangle$. The generalized eigenvectors can
now be found by solving the equation
$\left(E_H-d_0I\right)|\tilde{q}_r\rangle=|q_r\rangle$,
$\langle\tilde{q}_l|\left(E_H-d_0I\right)=\langle q_l|$. We next
define the matrices
$Q_l=\left(|q_l\rangle,|\tilde{q}_l\rangle\right)$,
$Q_r=\left(|q_r\rangle,|\tilde{q}_r\rangle\right)$ and the $2\times
2$ matrix $Q=\left(Q_l^TQ_r\right)^{-1}$. In the limit of large $N$,
it holds that
\[E_H^N\simeq Q_r\left(
             \begin{array}{cc}
               1 & N \\
               0 & 1 \\
             \end{array}
           \right)
           QQ_l^T.\]
As $Q_l^T|q_r\rangle=\left(
                       \begin{array}{c}
                         1 \\
                         0 \\
                       \end{array}
                     \right)=Q_r^T|q_l\rangle$,
the expectation value $\langle\psi|\hat{O}_N|\psi\rangle$ is given
by

\[\left(
    \begin{array}{cc}
      1 & N \\
    \end{array}
  \right) Q \left(
            \begin{array}{c}
              1 \\
              0 \\
            \end{array}
          \right)\]

in the limit of large $N$. The energy per site is therefore given by

\[Q_{12}=-1/\langle\tilde{q}_l|q_r\rangle=-1/\langle q_l|\tilde{q}_r\rangle=-1/\langle q_l|\left(E_H-d_0I\right)^{-\dagger}|q_r\rangle.\]

In a similar vein, it is possible to calculate expectation values of
the operator $\langle\psi|(H-\lambda)^2|\psi\rangle$. This is
relevant because it gives an exact bound on how far a given MPS
$|\psi\rangle$ is from an exact eigenvector of $H$. As we have a
squared term, Jordan blocks of dimension $3$ will be encountered. As
before, we define

\[E_{H^2}=\sum_{ijkl}A_i\otimes B_j\otimes \bar{B}_k \otimes \bar{A}_l \langle
l|X_k^\dagger X_j|i\rangle.\]

The relevant  right eigenvector is again of the form
$|q_r\rangle=|x_r\rangle|0\rangle|0\rangle$ and we can find the
eigenstructure of the associated Jordan block  as follows: start by
calculating $|\tilde{q}_r\rangle$ as we did in the previous section
using the operator $E_H$ instead of $E_{H^2}$; next define
$|\tilde{q}'\rangle={\rm sym}(|\tilde{q}_r\rangle|0\rangle)$ where
the symbol 'sym' means symmetrization with respect to the part of
the state acting on the Hamiltonian part of the MPO (the
antisymmetrized wavefunction turns out to be an irrelevant
eigenvector of $E_{H^2}$ with eigenvalue $d_0$); finally solve the
linear set of equations $\left(E_{H^2}-d_0
I\right)|\tilde{q}"_r\rangle=|\tilde{q}'_r\rangle$. The relevant
eigenstructure is now given by the matrix
$Q_r=\left(|q_r\rangle,|\tilde{q}'_r\rangle,|\tilde{q}"_r\rangle\right)$.
Similarly, one can calculate $Q_l$ and $Q=(Q_l^TQ_r)^{-1}$. The
final expectation value is then given by

\[\left(
    \begin{array}{ccc}
      1 ,& N ,& N(N-1)/2\\
    \end{array}
  \right) Q \left(
            \begin{array}{c}
              1 \\
              0 \\
              0 \\
            \end{array}
          \right)\]

The matrix $Q$ therefore contains all the relevant information about
the energies and their scaling when $N\rightarrow\infty$.

The energy can now easily be optimized with a brute force
gradient-base optimization routine.

Concerning the 2-dimensional MPO representing Hamiltonians, it turns
out that they are very valuable for speeding up actual calculations
done by the PEPS method: the calculation of the expectation value of
the Hamiltonian with respect to to a given PEPS can be calculated in
one run using this idea, and we don't have to calculate the
expectation value for every terms individually anymore.

\section{Conclusion}
In conclusion, we constructed several examples of interesting
families of matrix product operators in 1 and 2 dimensions. Those
descriptions turn out to be very valuable for constructing stable
and scalable algorithms for simulating quantum spin systems,  in one
and two dimensions.

\section{appendix}

In this appendix, we show how to solve the problem of approximating
any function $f(k)$ as a sum of exponentials for $k=1..N$:

\[\min_{x_i,\lambda_i} \sum_{k=1}^N|f(k)-\sum_{i=1}^n x_i\lambda_i^k|.\]

First, construct the rectangular $N-n+1\times n$ matrix

\begin{eqnarray*}F&=&\left(
      \begin{array}{ccccc}
        f(1) & f(2) & f(3)&\hdots & f(n) \\
        f(2) &  f(3) & \hdots & &  \\
        f(3) &  \hdots & & &  \\
        \vdots & &  &  & f(N-1) \\
        f(N-n+1) & \hdots & & f(N-1) & f(N) \\
      \end{array}
    \right)\\
    &\simeq&\underbrace{\left(
              \begin{array}{ccccc}
                \lambda_1^0 & \lambda_2^0 & \hdots & \lambda_n^0 \\
                \lambda_1^1 & \lambda_2^1 &  &    \\
                \lambda_1^2 &  &  &    \\
                 \vdots &   &   &    \\
                \lambda_1^{N-n} & \lambda_2^{N-n} &  \hdots & \lambda_n^{N-n} \\
              \end{array}
            \right)}_{W}
\left(
  \begin{array}{cccc}
    x_1 & 0 & \hdots & 0 \\
    0 & x_2 &  &  \\
    \vdots &  &  &  \\
    0 &  &  & x_n \\
  \end{array}
\right) \left(
  \begin{array}{cccc}
    \lambda_1^0 & \lambda_1^1 & \hdots & \lambda_1^n \\
    \lambda_1^0 & \lambda_2^1 & \hdots & \lambda_2^n \\
    \hdots &  &  &  \\
    \lambda_n^0 & \hdots &  &  \\
  \end{array}
\right)
\end{eqnarray*}
Note that $W$ is a Vandermonde matrix. We observe that $F$ and $W$
span the same space (note that $N$ is typically much larger than
$n$), such that there exists a $n\times n$ matrix $Q$ s.t. $FQ\simeq
W$. Define $F_1$ as the rectangular matrix which consists of the
first $N-n$ rows of $F$ and $F_2$ as the one with the last $N-n$
rows. Due to the Vandermonde structure of $W$, it must be
approximately true that $F_1Q\Lambda\simeq F_2Q$ with $\Lambda$ the
diagonal matrix containing the exponents. Therefore, $\Lambda\simeq
Q^{-1}F_1^{-\dagger}F_2Q$ ($F_1^{-\dagger}$ denotes the
pseudoinverse of $U_1$): the exponents $\{\lambda_i\}$ hence
correspond to the eigenvalues of the matrix $F_1^{-\dagger}F_2$
which can be calculated very easily.

This method can be made more robust by making use of the so--called
QR-decomposition. This can be done by first calculating the
(economical) QR decomposition of $F=UV$ and by defining $U_1$ as the
rectangular matrix which consists of the first $N-n$ rows and $n$
columns of $U$ and $U_2$ as the one with the last $N-n$ rows: there
must again exist a $Q$ such that $UQ\simeq W$. The exponents
$\Lambda$ can therefore easily be calculated as the eigenvalues of
the matrix $U_1^{-\dagger}U_2$. The advantage of using the
QR-decomposition is that the pseudoinverse of $U_1$ is much better
conditioned than of $F_1$.

Once those exponents are found, a simple least squares algorithm can
be used to find the corresponding weights $\{x_i\}$. It happens that
this method is very efficient and reliable, even when oscillating
functions are involved. A similar method is known in the field of
signal processing under the name of Hankel singular value
decomposition.

\end{document}